# Optimal piezo-electro-mechanical coupling to control plate vibrations

Silvio Alessandroni[1], Francesco dell'Isola[2], and Fabrizio Frezza[3]

[1]*Dott. in Meccanica Teorica ed Applicata, Università di Roma "La Sapienza", 00184 Roma, Italia*

[1]*Department of Engineering Science and Mechanics, Virginia Polytechnic Institute and State University, Blacksburg, VA 24061-0219, U.S.A.*

[2]*Dip. Ing. Strutturale e Geotecnica, Università di Roma "La Sapienza", 00184 Roma, Italy*

[3]*Dip Ing. Elettronica, Università di Roma "La Sapienza", 00184 Roma, Italy*

**Abstract**. A new way of coupling electrical and mechanical waves, using piezoelectric effect, is presented here. Since the energy exchange between two systems supporting wave propagation is maximum when their evolution is governed by similar equations, hence, an optimal electro-mechanical coupling is obtained by designing an electric network which is "analog" to the mechanical structure to be controlled. In this paper, we exploit this idea to enhance the coupling, between a Kirchhoff-Love plate and one possible synthesis of its circuital analog, as obtained by means of a set of piezoelectric actuators uniformly distributed upon the plate. It is shown how this approach allows for an optimal energy exchange between the mechanic and the electric forms independent of the modal evolution of the structure. Moreover, we show how an efficient electric dissipation of the mechanical energy can be obtained adding dissipative elements in the electric network.

**1. Introduction**

It is known that a deformation field of a piezoelectric material generates an electric field in it and vice versa (see [1]). However, it is not possible to obtain an efficient energy exchange between the electrical and mechanical waves propagating in a macroscopic structure of piezoelectric material, because the propagating speed of an electric field, in such a material (i.e., light speed) is much larger than the propagating speed of any deformation field (i.e., sound speed) [2]. This means that the electric field, generated by the deformation in a piezoelectric (PZT) material, is not supported and efficiently transmitted by the material itself, so that the energy be efficiently transformed from one form to the other (see [3]). Since light speed can be considered infinite when compared to sound speed, we can accept the lumped hypotheses for the used electrical components and we synthesize, in section 1, an electric circuit, the equations of which are analogous to the finite-difference equations for the deflection field of a plate. These circuits are called 'circuital analogs' of the considered mechanical systems [4]. They were used as analogical computers before the incoming of the digital ones for the design of mechanical structures (see [5]). In section 2 the interconnection between the plate and the circuital analog using piezoelectric actuators is considered and the model of the coupled electromechanical system is developed by means of a suitable homogenization procedure. We explicitly remark that no mathematical convergence result is obtained or used there. The identification procedure, which we developed, is purely heuristic and may deserve a more careful mathematical analysis similar to that developed in [6], [7]. In section 3, the analysis of the coupled system will show that the maximum energy transfer from the mechanical to the electrical form is independent of the modal evolution of the plate. Finally, considering the presence of dissipative elements in the analog, an efficient mechanical damping is obtained.

---

[1] Silvio Alessandroni: salessan@vt.edu
[2] Corresponding author: Francesco dell'Isola: francesco.dellisola@uniroma1.it
[3] Fabrizio Frezza: frezza@die.ing.uniroma1.it

**2. Electric analog synthesis**

Let us consider the Kirchhoff-Love plate dimensionless equation:

$$\nabla^2 \nabla^2 u + \alpha \ddot{u} = 0, \qquad \alpha = \frac{3\rho l_0^4}{h^2 E t_0^2} \qquad \forall\, u \in D \tag{1}$$

where $u$ and $\nabla^2 \nabla^2$ are the dimensionless deflection field and the double-Laplacian operator defined on the bidimentional domain $D$, the superscript dot represents the dimensionless time derivative, $l_0$ and $t_0$ are the characteristic length and time, $\rho$ is the plate mass density per unit volume, $h$ is the thickness of the plate, and E is the Young modulus of the material constituting the plate. Moreover, two different kinds of boundary conditions (BC) are chosen [2]:

- Completely-clumped BC:

$$\begin{cases} u|_{\partial D} = 0 \\ \frac{\partial u}{\partial n}\big|_{\partial D} = 0 \end{cases} \tag{2}$$

Where **n** is the normal to $\partial D$ in the tangent plane to $D$.

- Simply-supported BC:

$$\begin{cases} u|_{\partial D} = 0 \\ \frac{\partial^2 u}{\partial n^2}\big|_{\partial D} = 0 \end{cases} \tag{3}$$

The electric analog is found sampling the domain $D$ by a uniform-step grid and associating the electric tension of each node of the circuital analog to the deflection-velocity field in the nodes of the grid. Indeed let us introduce the Laplace transform of the field $u$. Therefore (1) becomes:

$$\nabla^2 \nabla^2 u + \alpha s^2 u = 0, \tag{4}$$

Where **s** is the Laplace variable. Then we replace the spatial differential operator $\nabla^2\nabla^2$ with its centred finite difference approximation. This means that we need to synthesize a lumped circuit in which the nodes correspond to the nodes of the grid and the Kirchhoff laws in terms of their voltage drop referred to ground become

$$\left(\frac{16}{s\varepsilon^4} + \alpha s\right)V[i,j] + \frac{1}{s\varepsilon^4}(V[i+2,j] - 6V[i+1,j] - 6V[i-1,j] + V[i-2,j]) +$$
$$+ \frac{1}{s\varepsilon^4}(V[i,j+2] - 6V[i,j+1] - 6V[i,j-1] + V[i,j-2]) + \quad (5)$$
$$+ \frac{1}{s\varepsilon^4}(V[i+1,j+1] - 6V[i-1,j+1] - 6V[i+1,j-1] + V[i-1,j-1]) = 0$$

where V is the dimensionless L-transform of the voltage drop $v$, $\alpha$ has to be identified in terms of electric immittances and $\varepsilon$ is the dimensionless sampling step used.

The evolution equations (5) present a coupling among the $ij$ node and its adjacent ones according to the figure (1)

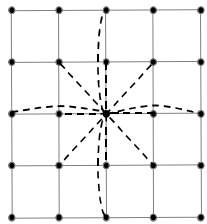

Figure 1: Involved nodes for the equation at i j node (dashed lines).

This implies the existence of a non-zero admittance edge in the electric analog between the ιφ node and all those to which it is connected. Thus, we can consider the topology in figure (2) for the circuital element of the analog limiting ourselves to the use of two-terminal networks (TTN) depicted as block boxes.

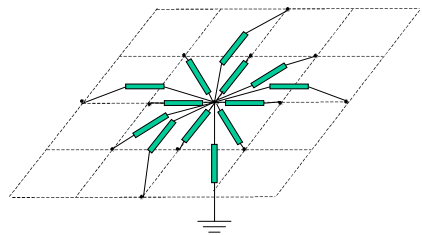

Figure 2: Circuit-elements topology

The balance equations for the current I at node $ij$ of the circuit in figure (2) are the following:

$$\sum_{m=-2}^{2}(I[i+m,j] + I[i,j+m]) + \sum_{n=-1}^{1}\sum_{m=-2}^{2} I[i+k,j+k] + I_G[i,j] = 0 \quad (6)$$

where $I_G(i,j)$ is the current flowing from the node $ij$ to the ground. The analog is synthesized assuming the connections between the electrical nodes constituted only by TTNs. Therefore, using the general expressions for the constitutive relation of a linear TTN, for each edge we obtain

$$I[i+p, j+q] = Y[i,j; i+p, j+q]V([i+p, j+q] - V[i,j]) \quad p,q = -2,..2 \tag{7}$$

where $Y[i,j; i+q, j+q]$ are the edge electric admittances. The values for the admittances are finally obtained equating the coefficients of (5) to those of (6) once transformed by using (7). In this way, the element of a possible electric analog of Kirchhoff-Love plate can be synthesized as shown by figure 3

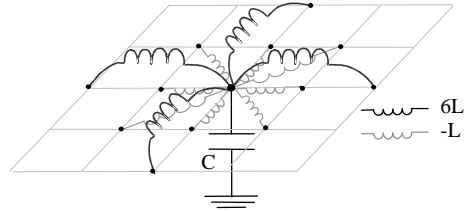

Figure 3: Analog circuital element

Where L and C are given by

$$L = \alpha \varepsilon^4 R_0 t_0 \qquad C = \frac{t_0}{R_0} \tag{8}$$

and $R_0$ is a characteristic dimensionless resistance. The presence of negative inductances (see figure (3)) is due to the assumption that the electric edges are constituted only by two-terminal networks and implies the presence of active elements able in their synthesis. Indeed, it is easy to realize negative admittances using the Antoniou circuit [8], the scheme of which is presented in figure (4)

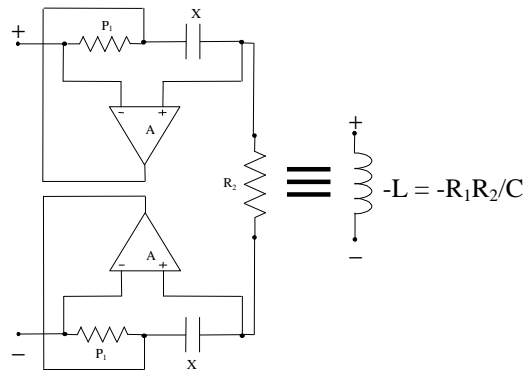

Figure 4: Negative inductance

In order to analyze the coupled system introduced in the next section, it is useful to consider the homogenized form of the circuital equations dependent on the time integral of the electric potential denoted by ψ:

$$\nabla^2\nabla^2\psi + \alpha\ddot{\psi} = 0, \qquad \alpha = \frac{LC}{t_0^2\varepsilon^4} \qquad \forall \psi \epsilon D \tag{9}$$

The previos equation allows us to state that the circuit defined by figure 3 is electric analog of Kirchhoff-Love plate if

$$LC = \frac{3\rho}{h^2 E} l_0^4 \varepsilon^4 \tag{10}$$

Let us explicitly remark that, once mechanical properties of the plate and the grid step are given, the condition (10) determines the value of the product $LC$.

**3. Piezoelectro-mechanical plate**

The electro-mechanical coupling between the plate and its circuital analog is realized using square piezoelectric actuators of negligible thickness which are centered at the sampling nodes of the plate (see (2)) and interconnect the nodes of the electric analog to the ground. Indeed, as electric components, they can be regarded as TTN of mainly capacitive admittance.

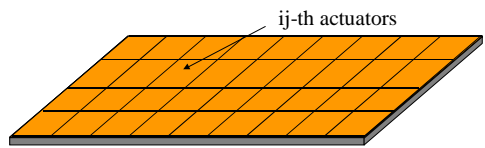

Fig. 5: Plate with actuators

Since the plate is described using a homogenized model while the circuit is a lumped device, it is convenient to describe the behavior of PZT actuators, which interconnect the two system, by the following constitutive relations, in which mechanical distributed variables, assumed constant on the domain of the actuator, and lumped electric quantities are used:

$$\begin{bmatrix} M_{a_{xx}} \\ M_{a_{yy}} \\ Q \end{bmatrix} = \begin{bmatrix} g_{mm} & g_{12} & -\frac{g_{em}}{b} \\ g_{21} & g_{mm} & -\frac{g_{em}}{b} \\ bg_{em} & bg_{em} & g_{ee} \end{bmatrix} \begin{bmatrix} u_{,xx} \\ u_{,yy} \\ v \end{bmatrix} \tag{11}$$

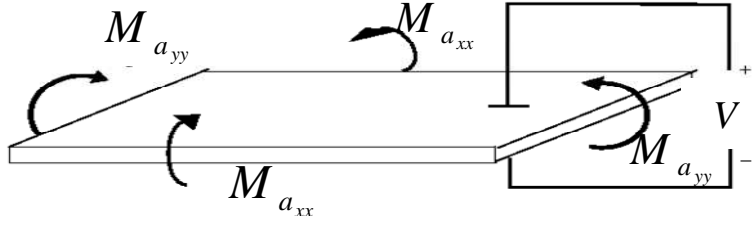

Figure 5: Bending actuator

where the bending-moment tensor components $M_{a_{xx}}, M_{a_{yy}}$ and the electric charge $Q$ stored into the actuator are related the plate curvatures $u_{,xx}, u_{,yy}$ and the voltage across the actuator terminals by the characteristic actuator coefficients $g$. The constant $b$ represents the edge length of the actuator.

In order to characterize the actuator from an electric point of view, let us consider the time derivative of the third relation of (11)

$$i = g_{ee}\dot{v} + sbg_{em}\nabla^2 \dot{u} \qquad (12)$$

where $i$ represents the current flowing in the actuator. Equation (12) shows how the actuator behaves like a capacitance in parallel connection with a current generator, the impressed value of which is related to the curvature of the actuator. Therefore, the use of the actuators as the capacitances of the electric analog is fully justified. The additional impressed currents modify (9) adding an electro-mechanical coupling term. In conclusion, the electric-coupled equation for the piezoelectro-mechanical (PEM) plate is

$$\nabla^2\nabla^2\psi + \alpha\ddot{\psi} + \beta_m \nabla^2 \dot{u} = 0 \qquad \beta_e = \frac{Lbg_{em}}{\varepsilon^4}\frac{l_0^7}{v_0 t_0^2} \qquad (13)$$

where $v_0$ is a characteristic voltage drop.

From a mechanical point of view, it is easy to consider the presence of the actuator layer as an additional term to the bending moment tensor of the plate, which is given by the first two relations of (11). Neglecting the contribution due to the purely-mechanical bending-stiffness of the actuator, the plate equation (1) is modified by an additional coupling term and gives the mechanical coupled equation of the PEM plate

$$\nabla^2\nabla^2 u + \alpha\ddot{u} + \beta_m \nabla^2 \dot{\psi} = 0 \qquad \beta_m = \frac{3g_{em}}{2bh^3 E} v_0 \qquad (14)$$

In the coupled system, the value of the circuital capacitance equals the actuator purely electric coefficient $g_{ee}$; therefore, the analogy condition given in (10) becomes the following condition for the circuit inductance L:

$$L = \frac{\varepsilon^4}{l_0^4}\frac{3\rho}{h^2 g_{ee} E} \qquad (15)$$

Moreover the equality between the coupling coefficients β in (13) and (4) can be obtained setting the characteristic voltage $v_0$ as follows

$$v_0 = l_0 \frac{b}{l_0} \sqrt{\frac{2h\rho}{g_{ee}}} \tag{16}$$

Therefore, the PEM plate equations are:

$$\begin{cases} \nabla^2\nabla^2\psi + \alpha\ddot{\psi} + \beta\nabla^2\dot{u} = 0 \\ \nabla^2\nabla^2 u + \alpha\ddot{u} - \beta\nabla^2\dot{\psi} = 0 \end{cases}, \quad \alpha = \frac{3\rho l_0^4}{h^2 E t_0^2}, \quad \beta = \frac{3 l_0 g_{em}}{bh^3 E} \tag{17}$$

## 4. Vibration in PEM plate

In order to observe the coupling properties in the piezo-electro-mechanical plate, let us consider a monochromatic electro-mechanical wave propagating in it

$$u = A e^{j(k \cdot r - \omega t)}, \quad \psi = B e^{j(k \cdot r - \omega t)} \tag{18}$$

where κ and $r$ are the wave vector and the position vector, respectively, and $\omega$ is the angular frequency. Replacing (18) into the (17), the dispersion relations for the PEM plate are obtained:

$$v_{p_1} = \frac{\omega_1}{k} = \frac{\beta}{2\alpha}\left(\sqrt{1 + \frac{4\alpha}{\beta^2}} + 1\right), \quad v_{p_2} = \frac{\omega_2}{k} = \frac{\beta}{2\alpha}\left(\sqrt{1 + \frac{4\alpha}{\beta^2}} - 1\right) \tag{19}$$

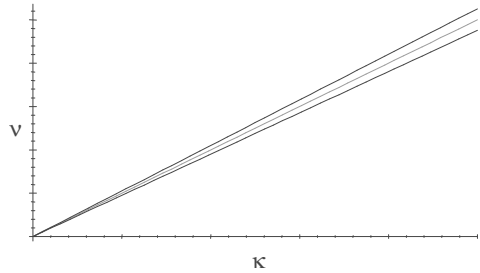

Figure 6: Dispersion relations

In figure (6), the phase speeds $v_p$ are plotted for the uncoupled case (gray line, β=0) and for the coupled one (black lines), in which it is apparent how the coupling generate a split of the dispersive relation around the uncoupled case.

Using the dispersion relations, it can be easily derived also the relation between the electric and mechanical wave-amplitudes A and B, which proves that the coupling is independent of the wave angular frequency and the wavelength:

$$A = iB \tag{20}$$

Therefore, the electric and the mechanical components show a phase difference of π/2 and are always equal in modulus. This means that, for monochromatic waves in PEM plates (as expected because of the results found in [9], [10], [11]) all the energy flows from one form to the other, thus realizing an optimal coupling.

Finally, a modal analysis is performed representing the solution $\{u, \psi\}$ on the basis $\{m_\lambda\}$ constituted by the eigenvectors of the problem (1) with simply supported boundary conditions:

$$u = \sum_\lambda p_\lambda(t) m_\lambda \qquad \psi = \sum_\lambda q_\lambda(t) m_\lambda \qquad (21)$$

where $p_\lambda(t)$ and $q_\lambda(t)$ are the mechanical and electric Fourier coefficients, respectively. The time evolution for any mode, is presented in figure (7) considering a purely mechanic deformation of the PEM plate at time t=0:

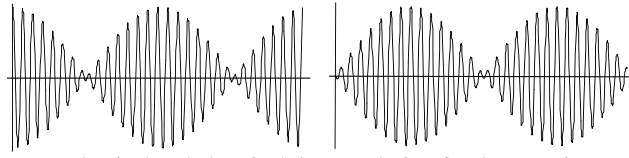

Fig. 7 Mechanical and electrical time-evolution for the generic e-m mode

Figure (7) proves again how the energy goes back and forth from the mechanic to the electrical forms.

If dissipative elements are introduced in the circuit-analog, then the mechanical energy converted into the electric form by the actuators can be damped. Hence, a non-vanishing purely resistive admittance in parallel connection with the actuator is added and the PEM plate relation (17) become

$$\begin{cases} \nabla^2\nabla^2\psi + \alpha\ddot{\psi} + \beta\nabla^2\dot{u} = 0 \\ \nabla^2\nabla^2 u + \alpha\ddot{u} + \beta\nabla^2\dot{\psi} + \gamma\dot{\psi} = 0 \end{cases}, \alpha = \frac{3\rho l_0^4}{h^2 E t_0^2}, \beta = \frac{3l_0 g_{em}}{bh^3 E}, \gamma = \frac{L}{t_0 R \varepsilon^4} \qquad (22)$$

where $R$ is the resistance.

Performing a modal analysis, as done for the non-dissipative case, an optimal resistance value can be found which is able to optimize the damping of the plate as shown in figure (8)

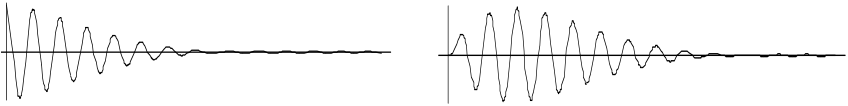

Fig. 8 Mechanical and electrical time-evolution for the generic e-m damped mode

The simulations have been performed considering an aluminum plate of 1 m of edge, and 1 mm of thickness with a grid of 10×10 commercial actuators, and an initial mechanical deformation of one per cent of the plate edge length. Let us remark that values for the circuital inductances of few Henry and voltages on the actuator less then 50 volts have been found, exhibiting the feasibility of the PEM plate.

## 5. Conclusions

This paper has demonstrated the possibility of obtaining an efficient vibration damping for a plate using a grid of piezoelectric actuators connected to the circuital analog of the plate. The approach used permits to obtain an optimum energy exchange and a vibration damping independent of the vibration mode of the plate. Moreover, the values found for the components used in the simulations show the technical feasibility of the PEM plate.

**Acknowledgements**

F. d. I. and S. A. wish to thank the Department of Engineering Science and Mechanics of the Virgina Polytechnic Institute and State University for the warm hospitality and research grants provided in the last two years.

**References**

[1] Eringen A.C., Maugin G.A. *Electrodynamics of Continua* I-II, New York Springer (1990)
[2] Courant R., Hilbert, D. *Methods of mathematical physics Vol. II Partial differential equations*. Interscience Publishers, New York London (1962).
[3] Bardati F., Barzilai G., Gerosa G., Elastic wave Excitation in PZT slabs, IEEE Transaction on sonics and ultrasonics, Vol. **SU-15**, n.4 October, (1968), 193-202.
[4] Molly C.T. *Four Pole Parameters in Vibration Analysis* in *Mechanical impedance Methods for Mechanical Vibrations* in *Colloquium on Mechanical Impedance Methods for Mechanical Vibrations* presented at the ASME Annual Meeting, New York, N.Y., December 2, (1958) sponsored by Shock And Vibrating Committee Applied Mechanics Division, The American Society of Mechanical Engineers, New York, edited by R.Plunkett.
[5] Kron G. *Equivalent circuits of the elastic field*, J. App, Mech. **11**, 149-161 (1944).
[6] Hoffmann, K. H., Botkin N. D. *Homogenization of Von Karman plates excited by piezoelctric patches*, ZAMM Z. Angew. Math. Mech. **89** (2000) n. 9, 579-590.
[7] Botkin N. D. Homogenization of an equation describing linear thin plates by piezopatches, Commun. Appl. Anal. 3 (1999) n. 2, 271-281.
[8] K.W.Wang *Structural vibration suppression via parametric control actions piezoelectric materials with real-time semi-active networks* in *Wave Motion, Intelligent Structures and Nonlinear Mechanics*, pp. 112-134 edited by A.Guran and D.J.Inman Series on Stability, Vibration and Control of Structures: Vol.1, World Scientific Singapore (1995).
[9] dell'Isola F., Vidoli S.: Bending-Waves Damping in Truss Beams by Electrical Transmission Line with PZT Actuators, Archive of Applied Mechanics **68**, 626-636, 1998
[10] Vidoli S., dell'Isola F.:Modal coupling in one-dimensional electro-mechanical structured continua, Acta Mechanica **141**, 1-2 (2000).
[11] Vidoli S., dell'Isola F. *Vibration control in plates by uniformly distributed actuators interconnected via electric networks*, European Journal of Mechanics/A, to appear.